\documentclass[11pt]{article}
\usepackage{amsmath,amsfonts,amssymb}
\usepackage{graphics}
\begin{document}

\newpage
\pagestyle{empty}

\vfill
  
\rightline{DSF-23/05}
\rightline{LAPTH-1108/05}
\rightline{q-bio.GN/0507030}
\rightline{July 2005}
 
 \vspace{20mm}

\begin{center}

{\Large \textbf{Conspiracy in bacterial genomes}}

\vspace{10mm}

{\large L. Frappat$^{ac}$, A. Sciarrino$^{b}$}

\vspace{10mm}

\emph{$^a$ Laboratoire d'Annecy-le-Vieux de Physique Th{\'e}orique LAPTH, 
CNRS, UMR 5108}

\emph{BP 110, F-74941 Annecy-le-Vieux Cedex, France}

\vspace{6mm}

\emph{$^b$ Dipartimento di Scienze Fisiche, Universit{\`a} di Napoli 
``Federico II"}

\emph{and I.N.F.N., Sezione di Napoli, Complesso Universitario di Monte S. 
Angelo}

\emph{Via Cintia, I-80126 Napoli, Italy}

\vspace{6mm}

\emph{$^c$ Member of Institut Universitaire de France}

\vspace{12mm}

\end{center}

\begin{abstract}
The rank ordered distribution of the codon usage frequencies for 109
eubacteria and 14 archaea is best fitted by a three-parameter function
which is the sum of a constant, an exponential and a linear term in the
rank $n$. The parameters depend (two parabolically) from the total GC
content. The rank ordered distribution of the amino acids is fitted by a
straight line. The Shannon entropy computed over all the codons is well
fitted by a parabola in the GC content, while the partial entropies
computed over subsets of the codons show peculiar different behavior.
Moreover the sum of the codon usage frequencies over particular sets, e.g.
with C and A (respectively G and U) as $i$-th nucleotide, shows a clear
linear dependence from the GC content, exhibiting a conspiracy effect.
\end{abstract}

\vfill

\begin{center}
  \textit{Revised version}
\end{center}

\vfill

\hrule

\vspace*{3mm}

\noindent
\emph{E-mail:} \texttt{frappat@lapp.in2p3.fr},
\texttt{sciarrino@na.infn.it}

\newpage
\pagestyle{plain}
\setcounter{page}{1}
\baselineskip=16pt



\section{Introduction}

The genetic code is degenerate, refering to the fact that almost all the
amino acids are encoded by `synonymous' codons, and this degeneracy is
primarily found in the third position of the codon. Some codons are used
much more frequently than others to encode a particular amino acid, the
pattern of codon usage varying between species. In the last few years, the
number of available data for coding sequences has considerably increased
\cite{NGI}, allowing for analysis to look for regularities, correlations
and general features over the whole exonic region. Recently, from an
analysis of the rank distribution for codons, in RNA coding sequences
\footnote{The DNA is constituted by four bases, the adenine (A), the
cytosine (C), the guanine (G) and the thymine (T), this last one being
replaced by the uracile (U) in the messenger RNA. A codon is defined as an
ordered sequence of three bases. Coding sequences in RNA are characterized
by their constituting codons.} performed in many genes for several
biological species, the existence of a universal, i.e. biological species
independent, distribution law for codons for the eukaryotic code has been
remarked \cite{FMSS}. Indeed, it was pointed out that the rank of codon
usage probabilities follows a universal law, the frequency function of the
rank ordered codons being very nicely fitted by a sum of an exponential, a
linear part and a constant. Such a universal behaviour suggested the
presence of general biases, one of which was identified with the total
\emph{exonic} GC content (denoted throughout the paper by
$y_{\mathrm{GC}}$, $0 \le y_{\mathrm{GC}} \le 1$), which is well known to
play a strong role in the evolutionary process. In fact, the values of the
parameters appearing in the fitting expression, plotted versus the total
exonic GC content of the biological species, were reasonably well fitted by
a parabolic function of the GC content. It is worthwhile to recall that the
determination of the kind of law the codon rank distribution follows is
extremely interesting in the investigations of the nature of the
evolutionary process, which has acted upon the codon distribution, i.e. the
eventual presence of a bias.

\medskip

Possible origins of codon bias have been recognized either as the result of
natural selection or as the result of mutational pressure acting on the
whole genome, also known as neutral evolution theory \cite{Kim68,KJ69}.
Experimental evidence has showed now that both effects should occur (see
e.g. \cite{MuOs87} for the influence of the biased mutation pressure in the
perspective of the neutral theory of molecular evolution, \cite{LSH02} for
exhibiting the role of external selective forces in the case of
thermophilic bacteria, \cite{CLHSMA} for an analysis of genome-wide codon
bias of bacterial species, \cite{SSPL93} for a discussion on the relative
role of mutational bias and translational selection for codon usage in many
genomes, including bacteria ones, \cite{KFL} for a model explaining trends
in codon and amino-acid usage vs. GC content). Quantitative model of
directional mutational pressure has been proposed to measure to some extent
the relative role of selection constraints and neutrality
\cite{Sueoka62,Sueoka88}. Many efforts have also been made to study codon
bias among genes of a given genome, in particular the influence of the gene
expression level \cite{GG82,ShLi86a,Bul91}, of the tRNA abundance
\cite{Ike85,ShLi86b}, of translational accuracy \cite{EW96}, see also
\cite{KFL} and references therein.

\medskip

It seemed to us interesting to continue our previous analysis on a `trend
across species' basis by focusing on the bacteria, one main interest of the
present study arising from the wide variation of the total GC content
ranging from 25\% to 75 \%, whereas the GC variation inside a bacterial
genome is much smaller. Our results is that the rank ordered distribution
of the codon usage frequencies for bacteria is best fitted by a three
parameters function, which is the sum of a constant, an exponential and a
linear term in the rank $n$. Two of the three parameters depend
parabolically on the total exonic GC content. As the sum, over suitable
sets below defined, of the codon usage frequencies is well fitted by a
straight line in the GC content, we say that \emph{conspiracy} effect has
to be present. Moreover, we conjecture the existence of an averaged
discrete symmetry in the codon usage frequencies, reasonably confirmed by
the data. We also calculate the rank ordered distribution of the 20 amino
acids, which is satisfactorily fitted by a straight line in
$y_{\mathrm{GC}}$.

We compute the Shannon entropy (as defined in \cite{Sha}) and find that its
behaviour in function of the exonic GC content is a parabola, whose apex is
around the value 0.50 of the GC content, which is expected for the
behaviour of the Shannon entropy for two variables. Moreover, the Shannon
entropies for the codons, whose orders in rank are, respectively, in the
ranges 1--15, 16--25 and 26--61, have peculiar features, which we comment
below.

The study of this paper was based on a sample of 109 bacteria and 14
archaea, with a codon statistics larger than 300\,000. The data were taken
from Codon Usage Tabulated from GenBank \cite{NGI} (see also
\texttt{http://www.kazusa.or.jp/codon/}), release 138 for eubacteria and
mainly release 144 for archaea.

\section{Codon usage probabilities distribution}

Let us define the usage probability for the codon $XZN$ ($X, Z, N \in
\{\mathrm{A, C, G, U}\}$)
\begin{equation}
  P(XZN) = \lim_{n_{tot} \to \infty} \;\;\; \frac{n_{XZN}}{N_{tot}} 
\label{eq:1}
\end{equation}
where $n_{XZN}$ is the number of times the codon $XZN$ has been used in all
considered processes, for a given biological species, and $N_{tot}$ is the
total number of codons used in the same processes. It follows that our
analysis and predictions hold for biological species with sufficiently
large statistics of codons. For each biological species, codons are ordered
following the decreasing order of the values of their usage probabilities,
i.e. codon number 1 corresponds to the highest value, codon number 2 is the
next highest, and so on. We denote by $f(n)$ the probability $P(XZN)$ of
finding $XZN$ in the $n$-th position. Of course the same codon occupies in
general two different positions in the rank distribution function for two
different species. By plotting $f(n)$ versus the rank we confirm that the
data are best fitted by the kind of function we found in \cite{FMSS}, i.e.
the sum of an exponential function, a linear function and a constant:
\begin{equation}
  f(n) = \alpha \, e^{-\eta n} \, - \, \beta \, n \, + \, \gamma \;.
  \label{eq:bf}
\end{equation}
In the following analysis we shall not consider the three stop codons,
whose function is very peculiar. Therefore the four parameters $\alpha$,
$\eta$, $\beta$ and $\gamma$ are constrained by the normalisation condition
\begin{equation}
  \xi = \frac{\alpha \, e^{-\eta}}{1 - e^{-\eta}} - 1891 \beta + 61 \gamma
  \label{eq:nor}
\end{equation}
where $\xi < 1$ is a number that is computed for any biological species
from the codon usage frequency for the 61 encoding codons (note that the
result is almost unchanged if the data are normalized to one summing over
the 64 coding codons). 
\footnote{The procedure to fit $f(n)$ in the present paper is slightly
different from the one followed in \cite{FMSS}. In that paper the value of
the parameter $\gamma$ was fixed to the value corresponding to a uniform
distribution, i.e. $\gamma = 1/61 \simeq 0.0164$. Therefore we were left
with only two free parameters, while in the present paper we use three free
parameters. While the general feature of the fit are unchanged, the present
fits are more accurate with a $\chi^{2}$ lower of about two orders of
magnitude.}

In the following, the parameters for the different fits have been computed
using a best-fit procedure, the curve fit being based on the
Levenberg-Marquardt algorithm \cite{numeric}. The $\chi^2$ coefficient is
defined by
\begin{equation}
  \chi^2 = \sum_{i} \frac{(y_{i}-y'_{i})^2}{y'_{i}}
  \label{eq:chi2}
\end{equation}
and the Pearson's $R$ coefficient by
\begin{equation}
  R = \frac{\displaystyle \sum_{i}
  (y_{i}-\overline{y})(y'_{i}-\overline{y}')}{\displaystyle \sqrt{\sum_{i}
  (y_{i}-\overline{y})^2}\sqrt{\sum_{i} (y'_{i}-\overline{y}')^2}}
  \label{eq:Rpearson}
\end{equation}
where $y_{i}$ are the actual values, $y'_{i}$ are the calculated ones,
$\overline{y}$ and $\overline{y}'$ are the means of the actual values and
of the calculated ones respectively. Recall that $R^2$ represents the ratio
of the explained variance on the total variance. Fits can be considered as
satisfactory for values of $R^2 \gtrsim 0.8$.

\begin{figure}[tbp]
  \centering
  \includegraphics{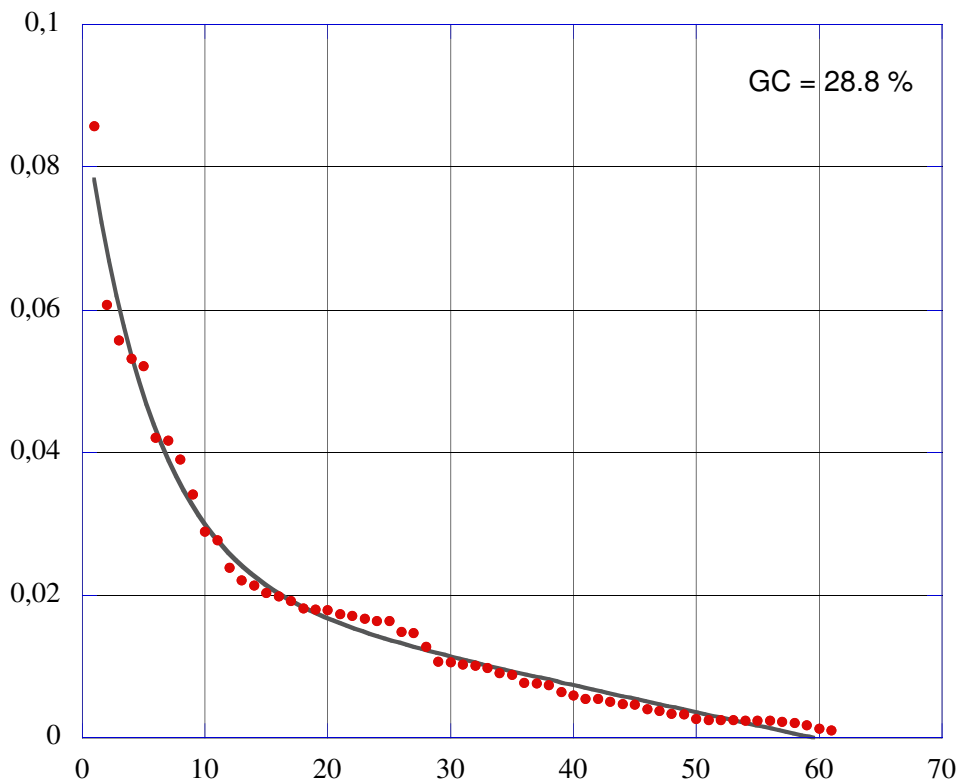}
  \caption{Codon rank distribution $f(n)$ for \textit{Borrelia burgdorferi}}
  \label{fig:fit1}
\end{figure}

\begin{figure}[tbp]
  \centering
  \includegraphics{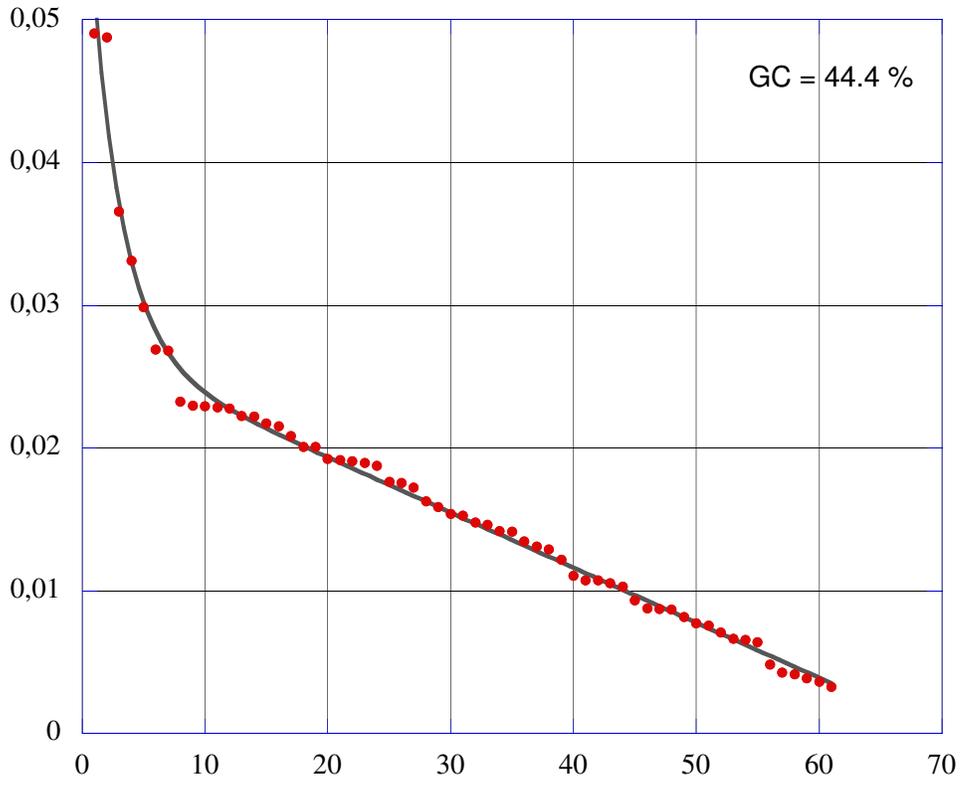}
  \caption{Codon rank distribution $f(n)$ for \textit{Bacillus subtilis}}
  \label{fig:fit2}
\end{figure}

\begin{figure}[tbp]
  \centering
  \includegraphics{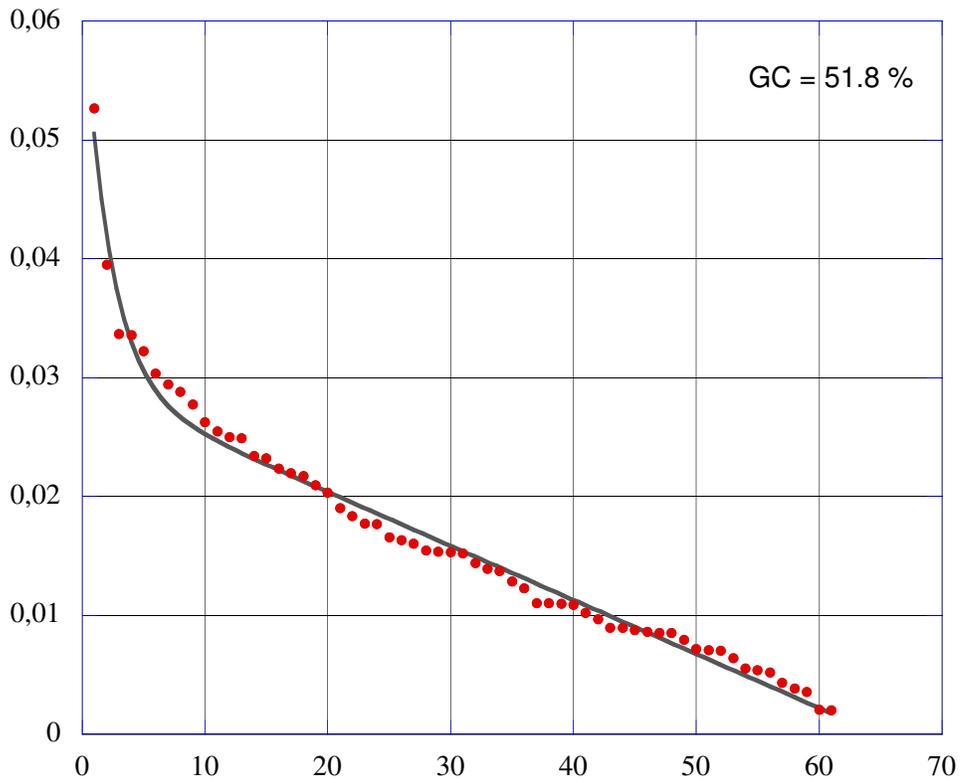}
  \caption{Codon rank distribution $f(n)$ for \textit{Escherichia coli}}
  \label{fig:fit3}
\end{figure}

\begin{figure}[tbp]
  \centering
  \includegraphics{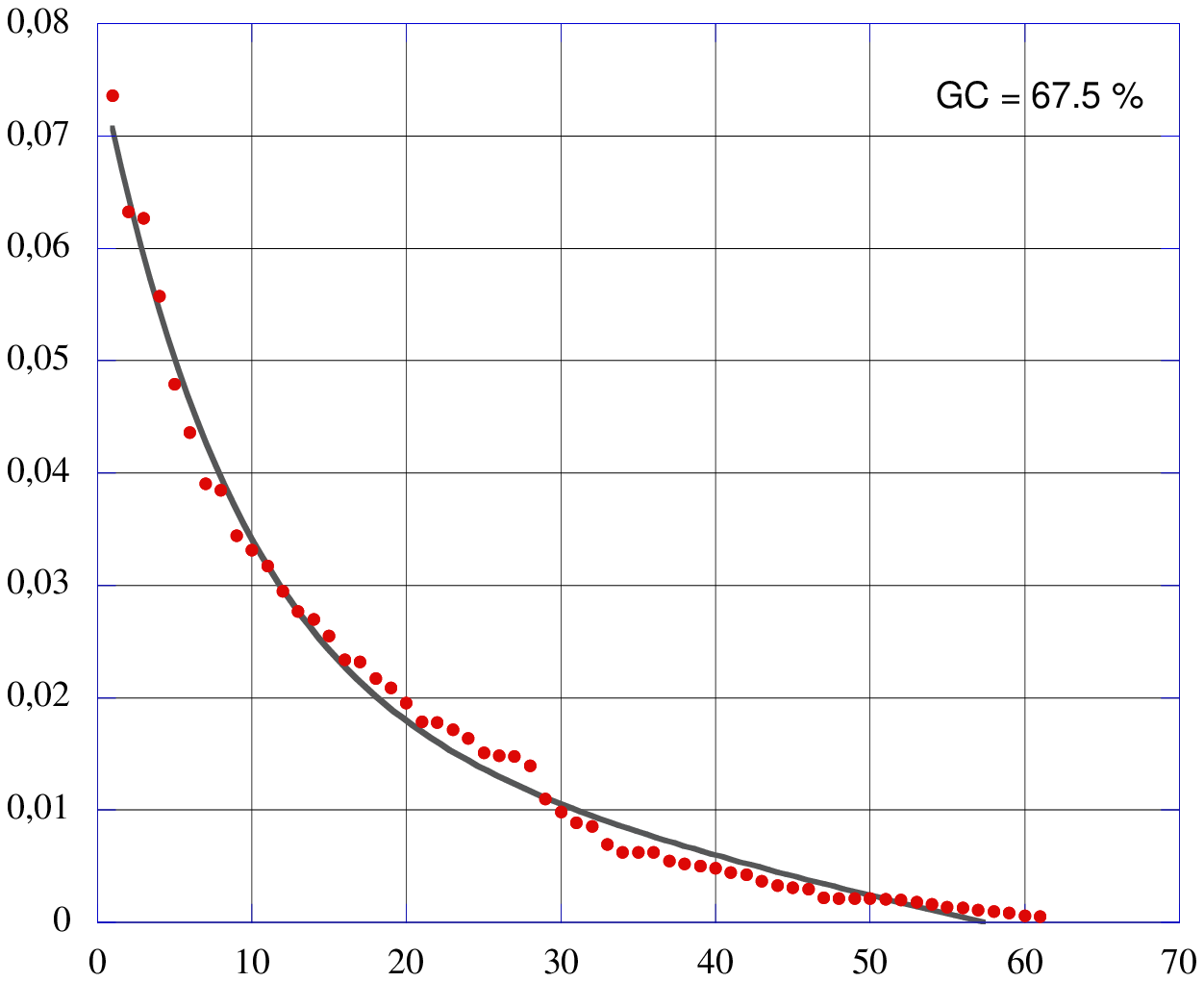}
  \caption{Codon rank distribution $f(n)$ for \textit{Ralstonia solanacearum}}
  \label{fig:fit4}
\end{figure}

In figs. \ref{fig:fit1}--\ref{fig:fit4}, we report the plot of $f(n)$ as a
function of $n$ for a few bacteria: \textit{Borrelia burgdorferi} (GC =
28.8 \%), \textit{Bacillus subtilis} (GC = 44.4 \%), \textit{Escherichia
coli} (GC = 51.8 \%), and \textit{Ralstonia solanacearum} (GC = 67.5 \%).
One remarks that the codon rank distributions show mainly a not far from
uniform or slowly decreasing behaviour for most of the codons, and a peak
of a small number of highly `overrepresented' codons (codon bias
contribution).
\footnote{Statistical tables specifying for each biological species the codon
statistics, the total GC content, the values of the parameters $\alpha$,
$\beta$, $\eta$, computed by a best-fit procedure, the estimated errors
$\Delta\alpha$, $\Delta\beta$, $\Delta\eta$ on these parameters, the
corresponding estimators of goodness of the fit $\chi^{2}$ and $R^2$ can be
found on the archives at the following address:
\texttt{http://xxx.lanl.gov/abs/q-bio.GN/0507030v1}.}


\begin{figure*}
  \centering
  \includegraphics{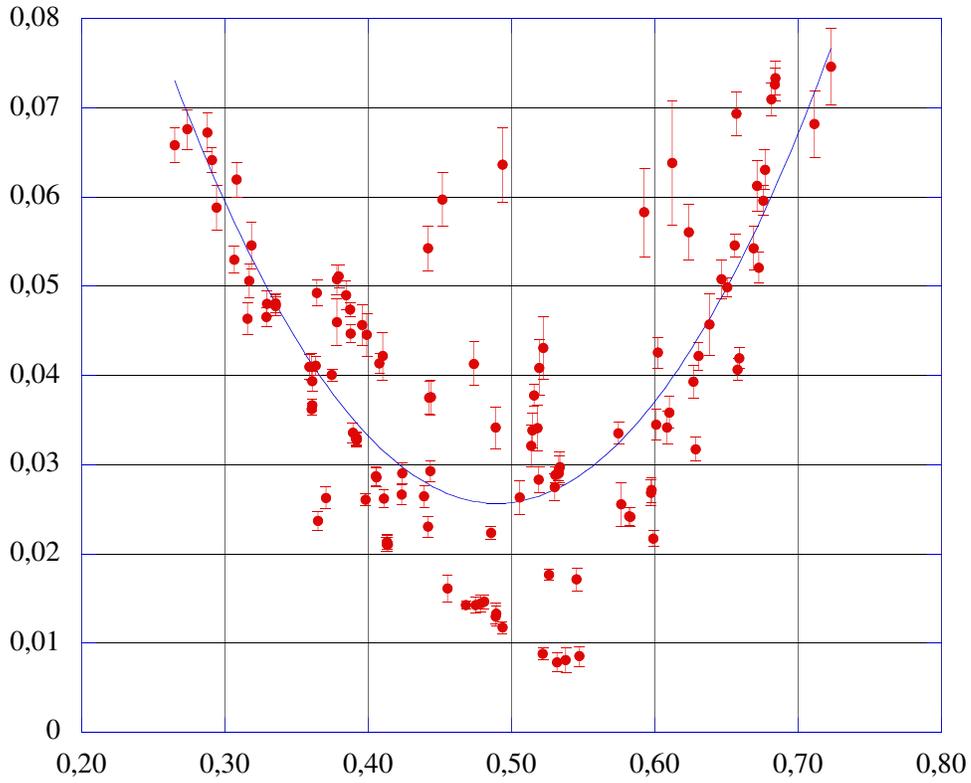}
  \caption{Parameter $\alpha$ vs. GC content}
  \label{fig:alphaGC}
\end{figure*}

\begin{figure*}
  \centering
  \includegraphics{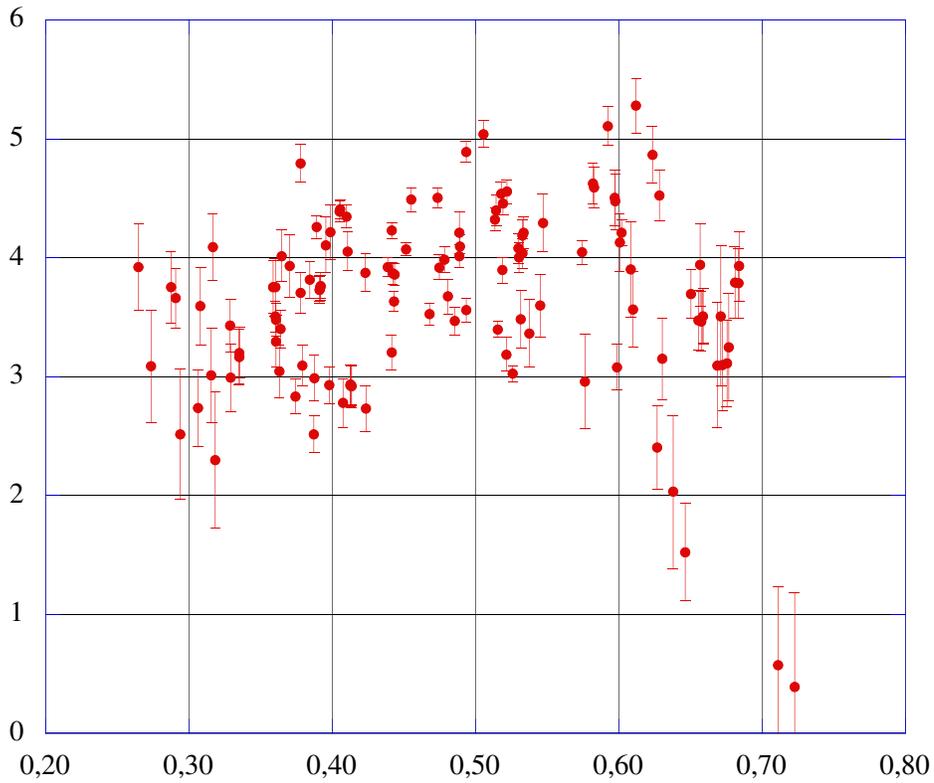}
  \caption{Parameter $10^{4} \, \beta$ vs. GC content}
  \label{fig:betaGC}
\end{figure*}

\begin{figure*}
  \centering
  \includegraphics{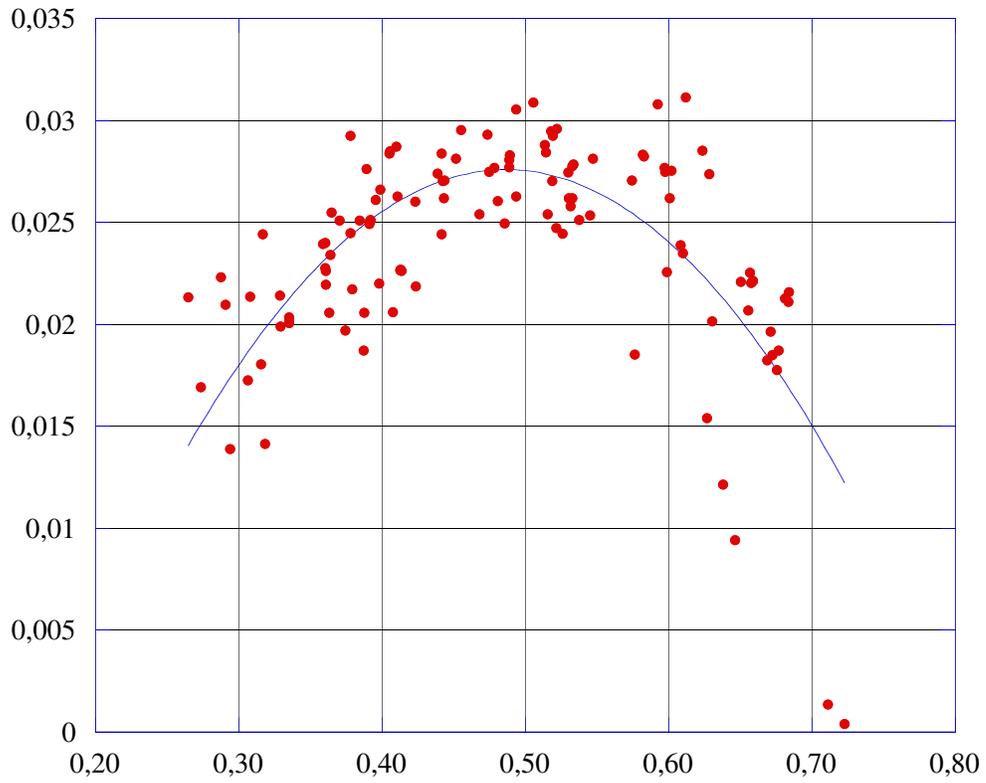}
  \caption{Parameter $\gamma$ vs. GC content}
  \label{fig:gammaGC}
\end{figure*}

\begin{figure*}
  \centering
  \includegraphics{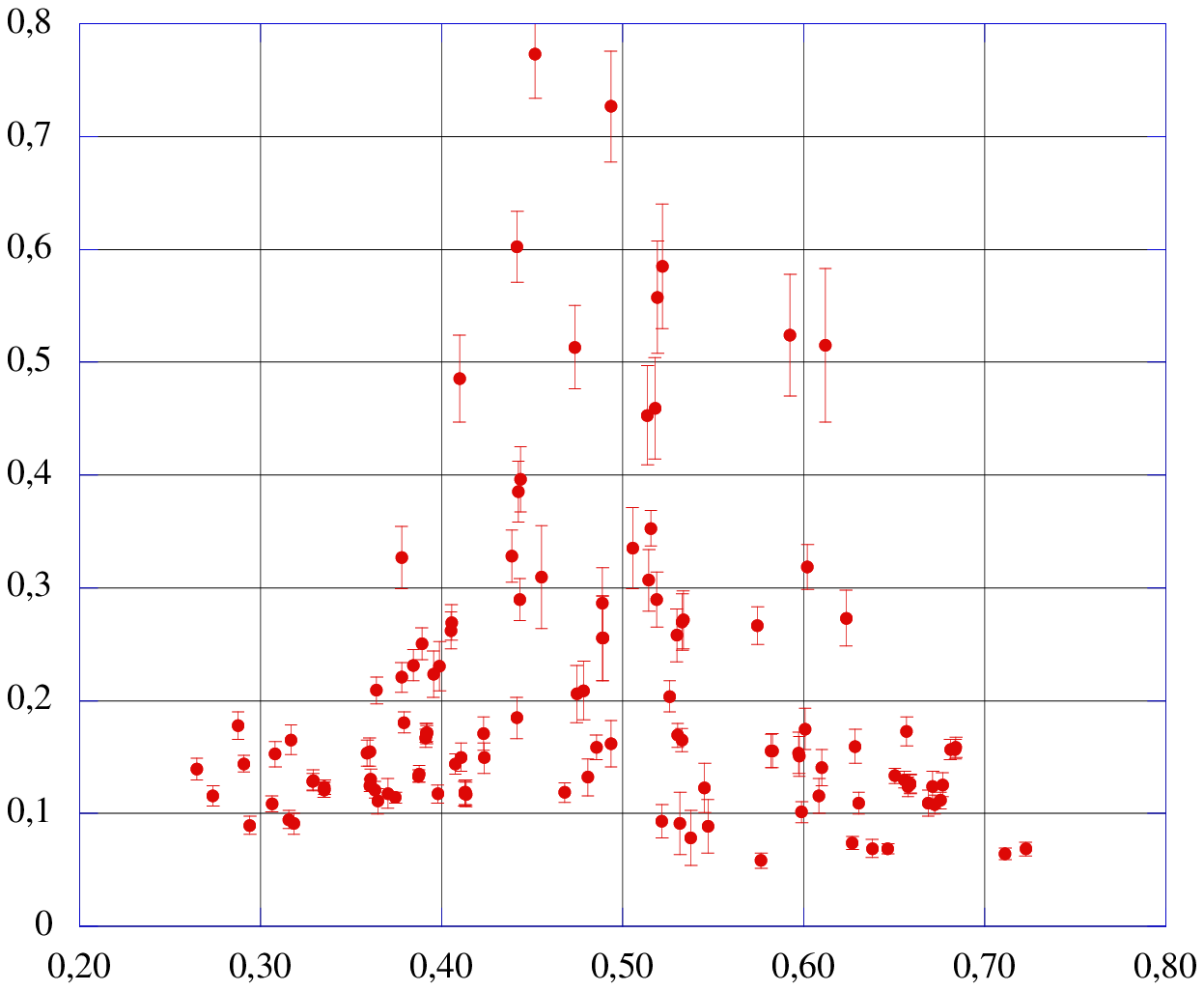}
  \caption{Parameter $\eta$ vs. GC content}
  \label{fig:etaGC}
\end{figure*}

Note that the fits of the rank ordered distribution $f(n)$ by a Yule law or
Zipf law are unsatisfactory. Indeed, as emphasized in \cite{MaKo96}, when
the majority of points resides in the tail of the distribution, it is
necessary to fit the whole range of data. From the previous discussion, we
expect the parameters to depend on the total GC content of the genes region
(here the total exonic GC content). We have investigated this dependence
and we report in figs. \ref{fig:alphaGC}--\ref{fig:etaGC} the plots of the
parameters $\alpha$, $\beta$, $\gamma$ and $\eta$ versus the total exonic
GC content. In terms of the total exonic GC content $y_{\mathrm{GC}}$ of
the biological species, one finds that the values of $\alpha$ and $\gamma$
are well fitted by polynomial functions:
\begin{equation}
  \alpha = \alpha_{0} + \alpha_{1} \, y_{\mathrm{GC}} + \alpha_{2} \,
  y_{\mathrm{GC}}^2
\end{equation}
where
\begin{eqnarray}
  \alpha_{0} = 0.250 \pm 0.017 \;, \quad \alpha_{1} = -0.919 \pm 0.071 \;,
  \quad \alpha_{2} = 0.939 \pm 0.072 \;,
  \label{eq:fitalpha}
\end{eqnarray}
the goodness of the fit being given by
\begin{equation}
  \chi^2 = 0.013 \;\;\mbox{and}\;\; R = 0.768 \;,
\end{equation}
and
\begin{equation}
  \gamma = \gamma_{0} + \gamma_{1} \, y_{\mathrm{GC}} + \gamma_{2} \,
  y_{\mathrm{GC}}^2
\end{equation}
where
\begin{equation}
  \gamma_{0} = -0.0376 \pm 0.0057 \;, \quad \gamma_{1} = 0.268 \pm 0.024
  \;,\quad \gamma_{2} = -0.275 \pm 0.024 \;,
  \label{eq:fitgamma}
\end{equation}
the goodness of the fit being given by
\begin{equation}
  \chi^2 = 0.0015 \;\;\mbox{and}\;\; R = 0.728 \;.
\end{equation}
The $\eta$ parameter shows a `$\lambda$-like' behaviour in terms of
$y_{\mathrm{GC}}$, centered on the value $y_{\mathrm{GC}} = 0.50$, while
the values of the $\beta$ parameter are mainly in the range $3~10^{-4} \le
\beta \le 5~10^{-4} $. Note that the errors on $\beta$ become large when
$y_{\mathrm{GC}}$ is far from the mean value 0.50.

Of course we are not able to predict which codon occupies the $n$-th rank.
Finally, let us remark that the total exonic GC content $y_{\mathrm{GC}}$
has to satisfy the consistency condition
\begin{equation}
  y_{\mathrm{GC}} =  \frac{1}{3} \, \sum_{i \in J} \, d_i f(i) \;,
  \label{eq:YGC}
\end{equation}
where the sum is over the set $J$ of integers to which the 56 codons
containing G and/or C nucleotides belong and $d_i$ is the multiplicity of
these nucleotides inside the $i$-th codon.

\section{Amino-acid rank distribution}

It is natural to wonder if some kind of universality is also present in the
rank distribution of amino acids. From the available data for codon usage,
we can immediately compute (using the bacterial code) the frequency of
appearance of any amino acid $F(n)$ ($1 \le n \le 20$) in the whole set of
coding sequences. The calculated values as a function of the rank are
satisfactorily fitted by a straight line,
\begin{equation}
  F(n) = F - Bn \;.
  \label{eq:FAA}
\end{equation}
In terms of the total exonic GC content, the parameters $B$ and $F$ for
eubacteria show a parabolic dependence: $B = B_{0} + B_{1} \,
y_{\mathrm{GC}} + B_{2} \, y_{\mathrm{GC}}^2$ where
\begin{equation}
  B_{0} = 0.00915 \pm 0.00034 \;, \quad B_{1} = -0.0232 \pm 0.0014 \;,
  \quad B_{2} = -0.0258 \pm 0.0014 \;,
  \label{eq:fitB}
\end{equation}
the goodness of the fit being given by
\begin{equation}
  \chi^2 = 4.4~10^{-6} \;\;\mbox{and}\;\; R = 0.911 \;,
\end{equation}
and $F = F_{0} + F_{1} \, y_{\mathrm{GC}} + F_{2} \, y_{\mathrm{GC}}^2$
where
\begin{equation}
  F_{0} = 0.145 \pm 0.004 \;, \quad F_{1} = -0.243 \pm 0.015 \;, \quad
  F_{2} = 0.270 \pm 0.015 \;,
  \label{eq:fitF}
\end{equation}
the goodness of the fit being given by
\begin{equation}
  \chi^2 = 48~10^{-5} \;\;\mbox{and}\;\; R = 0.911 \;.
\end{equation}
One remarks that the most frequent amino acid is always above the line.
This can be easily understood in the light of eq. (\ref{eq:bf}). Indeed,
the most frequent amino acids get, in general, a contribution of the
exponential term of (\ref{eq:bf}) with a low value of $n$.

\begin{figure*}
  \centering
  \includegraphics{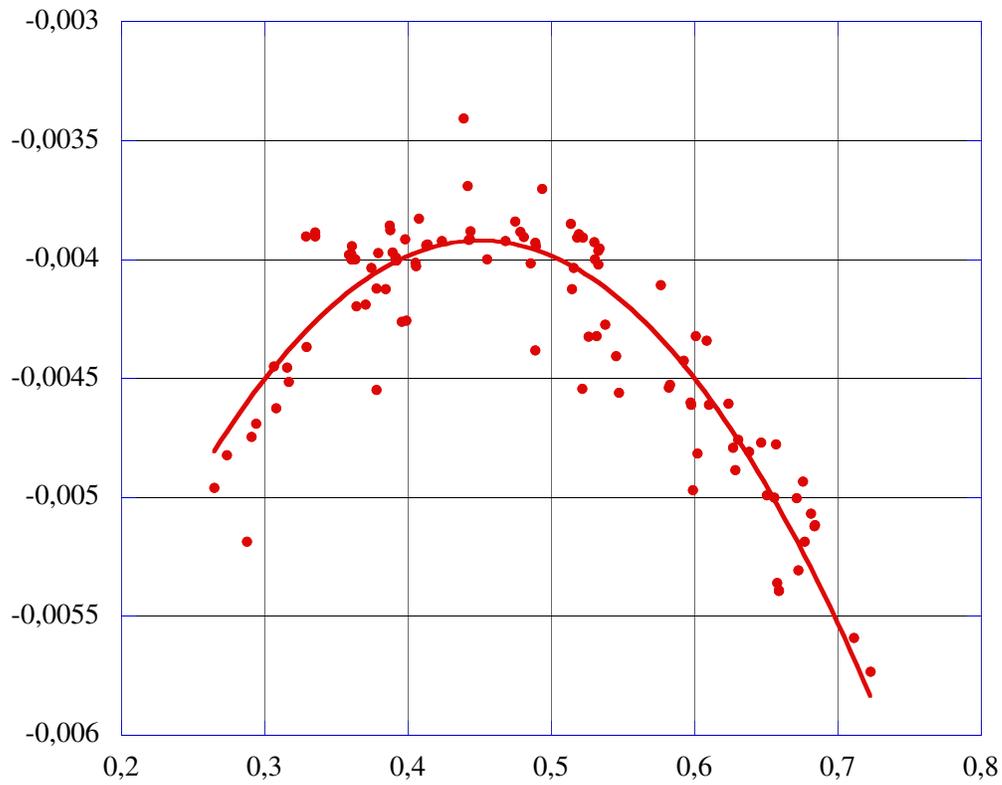}
  \caption{Parameter $B$ vs. GC content (eubacteria)}
  \label{fig:aminoBGC}
\end{figure*}

\begin{figure*}
  \centering
  \includegraphics{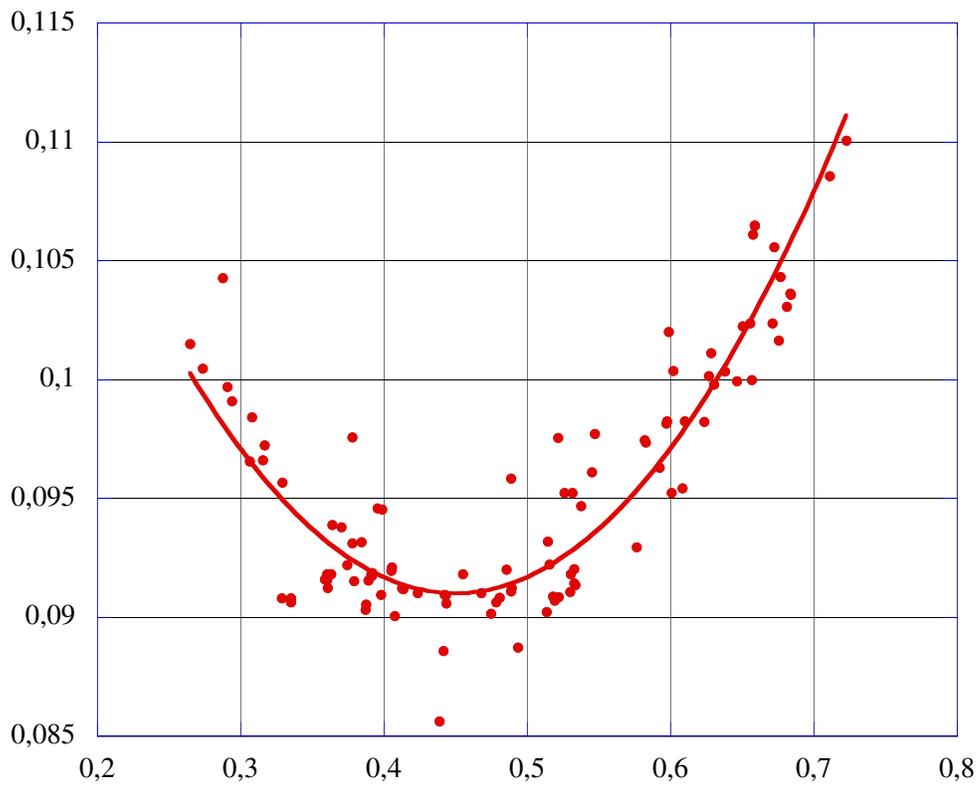}
  \caption{Parameter $F$ vs. GC content (eubacteria)}
  \label{fig:aminoFGC}
\end{figure*}

Of course, the frequency of an amino acid is correlated to frequencies of
its encoding codons given by (\ref{eq:bf}). If the ranks of the encoding
codons were completely random, we do not expect that their sum should take
equally spaced values, as is the case in a regression line. Therefore, we
can infer, for the biological species whose amino-acid frequencies are very
well fitted by a line, the existence of some functional constraints
(conspiracy effect) on the codon usage. An analysis of the influence of the
total GC content on the amino-acids composition of the protein of 59
bacteria has been reported in \cite{Lobry}, where references to previous
works on the subject can be found, see also \cite{Sueoka62, Sueoka88}. The
figures, reported in \cite{Lobry}, qualitatively agree with the linear
behavior given by eq. (\ref{eq:FAA}) and show sensible quantitative
differences with the expected frequencies, computed on the basis of the
neutral model, where the frequency of an amino acid is the sum of the
synonymous codon frequencies, the frequency of a codon being computed as
the product of the probability of the three nucleotides (mean field model).
This analysis already hints in the direction of the presence of functional
constraints.

However, the behaviour predicted by (\ref{eq:bf}) fits the experimental
data very well, while the shape of the distribution of amino acids seems
more sensible to the biological species. In fact, one can remark on many
plots of the amino-acid distributions the existence of one or two plateaux,
which obviously indicate an equal probabilities of use for some
amino acids. Presently, we have neither any arguments to explain the
uniform distribution of amino acids from the ranked distribution of the
corresponding codons nor we know of any explanation of this pattern of
distribution.

\section{The Shannon entropy}

\begin{figure*}
  \centering
  \includegraphics{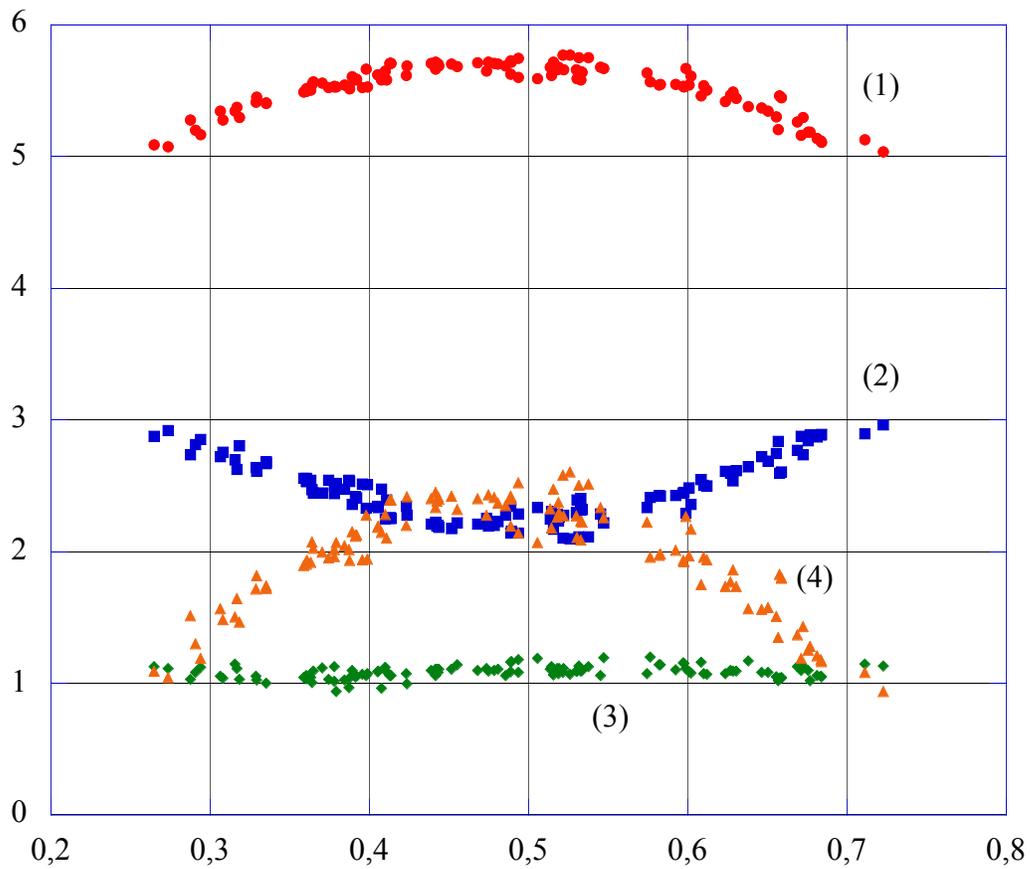}
  \caption{Shannon entropy and partial entropies vs. GC content: (1)
  total entropy, (2) partial entropy rank 1--15, (3) partial entropy
  rank 16--25, (4) partial entropy rank 26--61}
  \label{fig:pe}
\end{figure*}

We compute the Shannon entropy, 
\begin{equation}
  S = - \sum_{n=1}^{n=61} \, f(n) \, \log_2 f(n) \;,
\end{equation}
related to the codon rank distribution and plot it versus the total exonic
GC content $y_{\mathrm{GC}}$ for the sample of 109 eubacteria and 14
archaea, see fig. \ref{fig:pe}. The contribution of the stop codons is
negligeable and represents less than 0.5\% of the total entropy. The
Shannon entropy is very well fitted by a parabola,
\begin{equation}
  S = s_{0} + s_{1} \, y_{\mathrm{GC}} + s_{2} \, y_{\mathrm{GC}}^2
\end{equation}
where
\begin{equation}
  s_{0} = 2.670 \pm 0.084 \;, \quad s_{1} = 12.36 \pm 0.35 \;, \quad s_{2}
  = -12.73 \pm 0.36 \;,
  \label{eq:fitS}
\end{equation}
the goodness of the fit being given by
\begin{equation}
  \chi^2 = 0.329 \;\;\mbox{and}\;\; R^2 = 0,916 \;.
\end{equation}
Note that the parabola has its apex for $y_{\mathrm{GC}} \approx 0.50$,
which is expected for the behaviour of the Shannon entropy for two
variables (here GC and AU). We have computed the partial Shannon entropies
for the codons whose orders in rank are, respectively, in the ranges 1--15,
16--25 and 26--61 and we report them in fig. \ref{fig:pe} versus the GC
content. In any of the three sets, we have put the codons whose
contributions to the entropy, with respect the GC content, have similar
behaviour. Indeed, the codons of the first set are primarily influenced by
the exponential term in the rank distribution (\ref{eq:bf}), leading for
the partial entropy $S_{1-15}$ to a parabolic behaviour with positive
curvature and minimum at 50\% GC content. For the codons of the
intermediate set, the exponential term and the last two terms are of the
same order of magnitude, hence the partial entropy $S_{16-25}$ is almost
uniform with respect to the GC content. For the codons of the last set, the
exponential term is completely negligeable, and since $f(n)$ is very small,
$-f(n) \log_2 f(n) \simeq f(n)/\ln 2$. The trend of the partial entropy
$S_{26-61}$ is thus essentially given by the behaviour of the $\gamma$
parameter, hence the parabolic shape with negative curvature and maximum at
50\% GC content.

The fact that the Shannon entropy is a parabola shows obviously that the
codon distribution is not uniform: the uniform distribution corresponds to
the maximal entropy $S = \log_2 N$, independent of the GC content, where
$N$ is the number of considered codons (here $N=61$). The concavity of the
entropy is also well understood in this context in terms of deviations to
the uniform distribution when $y_{\mathrm{GC}}$ is far from the mean value
50\%. On the contrary, the partial entropy $S_{1-15}$, which is associated
to the `overrepresented' codons, exhibits a convex shape. This feature can
also be understood by realizing that in the case of $y_{\mathrm{GC}}$ far
from 50\%, the overrepresented codons distribution would be closer to the
expected value for a non biased distribution than in the case of
$y_{\mathrm{GC}} \approx 50\%$.

\section{The mystery of the straight lines}
  
In recent papers \cite{GPZ,GZ04}, it has ben remarked, for 129 eubacteria
and for 19 archaea, that the codon-specific nucleotide frequency
$P^{i}_{X}$, i.e. the sum of the codon usage for any given nucleotide $X$
in any fixed $i$-th position, that is, for example
\begin{equation}
  P^{1}_{X} = \sum_{Y,Z } \, P(XYZ) \;,
  \label{eq:csf}
\end{equation}
is fitted, versus the GC content, by a straight line, with coefficients
depending on the position and on the nature of the nucleotide and slightly
different for eubacteria and archaebacteria (see Table \ref{table:lineareu}
for eubacteria and Table \ref{table:lineararchae} for archaea for the
values of the slope and the axis intercept in the case of our sample of
bacteria). 
\footnote{Note that the difference of behaviour between eubacteria and
archaea could stem from the fact that eubacteria (at least in our sample)
are mainly mesophilic while archaea are essentially thermophilic.} 
There are 12 codon-specific nucleotide frequencies, which are constrained
by three normalisation conditions
\begin{equation}
  P^{i}_{C} + P^{i}_{U} + P^{i}_{G} + P^{i}_{A} = 1 \;, \qquad i = 1,2,3
  \;.
  \label{eq:ncsf}
\end{equation}
The fact that the sum of the codon position-specific nucleotide frequencies
is a linear function of GC is a mysterious feature, which becomes more
mysterious in the light of the results of Sec. 2, where we have remarked
that the parameters appearing in the rank ordered frequency (\ref{eq:bf})
are parabolic functions of GC. A \emph{conspiracy} has to be present
between the $f(n)$, see eq. (\ref{eq:bf}), for any bacteria such that the
sum of $f(n)$ over the 16-dimensional sets $Q$, depending on the considered
bacteria and corresponding the codon position-specific frequencies produces
a linear function of GC, that is
\begin{equation} 
  \sum_{n \in Q} \, f(n) = a \, y_{\mathrm{GC}} \, + \, b
  \label{eq:cf}
\end{equation}
Things become still more mysterious taking into accounts what we have
remarked, guided by the mathematical structure of the \emph{crystal basis
model} of the genetic code \cite{FSS}. Let us recall that in this model the
four nucleotides are assigned to the 4-dim fundamental irreducible
representation $(\mathbf{1/2, 1/2})$ of $ \mathcal{U}_{q \to 0}(so(4))
\equiv \mathcal{U}_{q \to 0}(sl(2) \oplus sl(2) = \mathcal{U}_{q \to
0}(sl_H(2) \oplus sl_V(2))$, (the lower labels $H$ and $V$ denote the two
commuting $sl(2)$), with the state assignment, (in the following the first
number denotes the value of the label $J$ specifying the irreducible
representation (irrep.) of $sl(2)$ and the second one the value of the
label $J_3$ denoting the state in the irrep., $J_3 = J, J-1, \ldots,-J$,
$2J \in \mathbb{Z}_+$)
\begin{eqnarray}
  \mathrm{C} & \equiv & (\mathbf{1/2, 1/2})_{H},(\mathbf{1/2, 1/2})_{V}
  \nonumber \\
  \mathrm{U} & \equiv & (\mathbf{1/2, -1/2})_{H},(\mathbf{1/2, 1/2})_{V}
  \nonumber \\
  \mathrm{G} & \equiv & (\mathbf{1/2, 1/2})_{H},(\mathbf{1/2, -1/2})_{V}
  \nonumber \\
  \mathrm{A} & \equiv & (\mathbf{1/2, -1/2})_{H},(\mathbf{1/2, -1/2})_{V}
\end{eqnarray}
which, in matrix notation, can be written as
\[
  \left(
  \begin{tabular}{cc}
    C & U \\
    G & A \\
  \end{tabular} 
  \right)
\] 
The codons are the composite states of the 3-fold tensor product of the
fundamental irreducible representation, that is
\[
  \left(
  \begin{tabular}{cc}
    C & U \\
    G & A \\
  \end{tabular} 
  \right)^{\otimes 3}
\] 
which can be written in matrix form as
\begin{equation}
	\left(
  	\begin{tabular}{cccc|cccc}
   	CCC & CCU & CUC & CUU \; & \; UCC & UCU & UUC & UUU \\
    CCG & CCA & CUG & CUA \; & \; UCG & UCA & UUG & UUA \\
    CGC & CGU & CAC & CAU \; & \; UGC & UGU & UAC & UAU \\
    CGG & CGA & CAG & CAA \; & \; UGG & UGA & UAG & UAA \\
    \hline
    GCC & GCU & GUC & GUU \; & \; ACC & ACU & AUC & AUU \\
    GCG & GCA & GUG & GUA \; & \; ACG & ACA & AUG & AUA \\
    GGC & GGU & GAC & GAU \; & \; AGC & AGU & AAC & AAU \\
    GGG & GGA & GAG & GAA \; & \; AGG & AGA & AAG & AAA \\
  \end{tabular}
  \right)
  \label{eq:codonmatrix}
\end{equation}
Eq. (\ref{eq:ncsf}) can be rewritten, for instance for $i = 1$, as
\begin{equation}
  P^{1}_{\mathrm{C}} + P^{1}_{\mathrm{A}} - 1/2 = - P^{1}_{\mathrm{U}} -
  P^{1}_{\mathrm{G}} + 1/2
\end{equation}
and, looking at the codon matrix, we remark that the l.h.s., respectively
the r.h.s., is, up to $\mp 1/2$, the sum of the two $4 \times 4$ matrices
obtained by \textbf{mirror inversion} with respect to the center of the
matrix, that is the exchange C $\to$ A, G $\to$ U. \footnote{The mirror
inversion in the crystal basis formalism means change of the sign of
$J_{3,H}$ and $J_{3,V}$.} So we make the following conjecture.
The sum of codon usage frequencies over a set of $r$ codons $(4 \le r \le
16)$, placed on the same side of the diagonals of the matrix, plus the
mirror set is described by a straight line in $y_{\mathrm{GC}} $ with a
very small slope coefficient $a$ ($|a| < 0.1$) and a constant term $b$
equal to $r/32$. In other words, the sum of codon usage frequencies over a
set of $r$ codons, placed on the same side of the diagonals of the matrix
with a plus the numerical factor $-r/16$, is equal, up a factor $r/32$, to
the opposite of the sum of the codon usage frequencies over the set of $r$
mirror codons. The mirror symmetry with respect to the secondary diagonal
(resp. principal diagonal) is equivalent, in terms of $\mathcal{U}_{q \to
0}(sl_H(2) \oplus sl_V(2)) $ to the change of the sign of the third
component of $J_{3,H}$ and $J_{3,V}$ (resp. of the sign of the third
component of $J_{3,V}$). Denoting $P^{i}(XYZ)$ (resp.
$P^{i}(\overline{XYZ})$) the codon usage frequency of the $i$-th codon
$XYZ$ (resp. of the mirror codon $\overline{XYZ})$, we make the ansatz that
the following equation holds
\begin{equation}
  \sum_{k \in I} \, \left (P^{k}(XYZ) + P^{k}(\overline{XYZ}) \right ) = a
  \, y_{\mathrm{GC}} + \, \frac{r}{32}
  \label{eq:form28}
\end{equation}
where $I$ is $r$-dimensional set of codons placed on the same side of the
principal diagonal or secondary diagonal of the codon matrix . The mirror
codon $\overline{XYZ}$ with respect to the secondary diagonal (resp. to the
principal diagonal) has opposite values of $J_{3,H}$ and $J_{3,V}$ of the
$XYZ$ codon (respectively, opposite values of $J_{3,V}$). One can see from
Table \ref{table:sort} that indeed the values of the axis intercept $b$ are
very close to the conjectured values $r/32$ (4/32 = 0.125, 8/32 = 0.250,
12/32 = 0.375).
So our conjecture seems supported by the experimental data and suggests the
existence of an {\it averaged symmetry} C $\to$ A, G $\to$ U or C $\to$ G,
U $\to$ A for the codon usage frequencies.
\footnote{These symmetries have already appeared in the literature in a
different context. Indeed Rumer \cite{Ru}, as quoted in \cite{Shch},
remarked in the sixties, that the first symmetry exchanges the 32
\emph{strong codons}, which form quartets or the quartet subset of sextets,
with the remaining 32 \emph{weak codons}. This symmetry is sometimes
referred to as the Rumer symmetry \cite{Shch}. The second symmetry has been
remarked by \cite{KoRu} as the symmetry which transforms each of the
\emph{octets} into itself, an \emph{octet} being a set of eight
dinucleotides appearing or not appearing as the first two nucleotides in
the quartets.}
Of course when $2r$ is equal to the number of the codons in the considered
set, the above conjecture is trivially the normalisation condition.

An analysis over our sample of 109 eubacteria, randomly drawing 4, 8, 12
codons in the 16-dim set of codons with a specified nucleotide in a fixed
position, shows that the straight line behaviour is surprisingly present
already for 4 codons (see Table \ref{table:sort}), where we present few
examples which confirm the conjecture.
In this table, $N$ denotes the number of random drawings of $n$ different
codons belonging to a given set $I$ (e.g. in the first line of Table
\ref{table:sort}, $I = \{ \mathrm{C}NN' \}$, $N,N' = \mathrm{A, C, G, U}$).
For each drawing lot of codons in the set $I$, we expect the sum
$\displaystyle \sum_{XYZ \in I} P(XYZ) + P(\overline{XYZ})$ to behave
linearly in terms of the GC content $y_{\mathrm{GC}}$, with slope $a$ and
axis intercept $b$. We observe that the distribution of the coefficients
$a$ and $b$ is peaked around mean values $\overline{a}$ and $\overline{b}$
with standard deviations $\sigma_{a}$ and $\sigma_{b}$, which are reported
in Table \ref{table:sort}. Two particular subsets have also been
considered, $I_{p} = \{ \mathrm{UNN}', \mathrm{CUN, CCC, CCU, CCA, CAC,
CAU, CAA, AUN, ACU, AAU} \} $ and $I_{s} = \{ \mathrm{CNN}', \mathrm{GCN,
GGC, GGG, GGU, GUC, GUG, GUU, UCN, UUC, UGC} \}$, corresponding
respectively to codons above the principal and secondary diagonals of the
codon matrix (\ref{eq:codonmatrix}). We have summed the experimental codon
usage of these codons and of the $n$ mirror codons with respect to the
secondary diagonal. As it can be read from Tables \ref{table:lineareu} and
\ref{table:sort}, no real meaningful difference appears between $n < 16$
and $n = 16$ for the first six sets $I$ of Table \ref{table:sort}, which
correspond to fix the C or G codons (hence the A or U codons) in first,
second and third position respectively. Note that the number of possible
different configurations is 1946 for $n = 4$ or 12 and 12870 for $n = 8$.
Surprisingly, this behavior is more evident for codons with fixed
nucleotides in \emph{third} position, that is for codons encoding different
amino acids, strongly suggesting strong constraints between amino acids.

Finally, let us remark that such a behaviour does not show up any longer if
one chooses for $I$ sets which violates the diagonal symmetry (for example
the first four lines or the first four columns of the codon matrix
(\ref{eq:codonmatrix}), or the whole set of codons) and considers random
drawings in the same conditions as above.

So, we derive the conclusion that the sum of the codon usage frequencies
over a suitable set of codons is a linear function of GC, showing a
\emph{conspiracy} effect between the $f(n)$ similar to the one of eq.
(\ref{eq:cf}), that is
\begin{equation} 
  \sum_{n \in Q} \, f(n) = a \, y_{\mathrm{GC}} \, + \, b
  \label{eq:cf2}
\end{equation} 
where now $Q$ is any $2r$-dimensional set of codons satisfying the above
discussed characteristics.

\medskip

The results above were based on sums of codons of the type $P(XYZ) +
P(\overline{XYZ})$, the bar meaning the \emph{mirror} codon obtained by the
mirror symmetry C $\leftrightarrow$ A, G $\leftrightarrow$ U. It is
interesting to see what happens if one considers instead sums of codons and
their \emph{reverse complementary} codons, i.e. $P(XYZ) + P(\widehat{ZYX})$
where the hat means the complementary rule C $\leftrightarrow$ G, A
$\leftrightarrow$ U (in other words, summing codons on one strand and
complementary codons on the other strand). A preliminary analysis shows
that the behaviour (\ref{eq:form28}) also holds, but the slope and axis
intercept ranges are wider and the mean slope is not close to zero. Such an
analysis could be instructive because it might shed some light on the DNA
strand asymmetry \cite{Sueoka95, Lobry96, MrKa98, Kow01}. However, in the
present work, we analyse the coding sequences extracted from the CUTG
database, for which one does not known whether the coding sequence comes
from a gene on the leading strand or on the lagging strand, and the
intergenic (non coding) sequences were obviously not considered. Hence we
are unable to comment on this point for the moment.

\section{Discussion and conclusion}

The distribution of the experimental codon probabilities for the total
exonic region for a large sample of bacteria is well fitted by the law
(\ref{eq:bf}). The spectrum of the distribution is universal, but the
codon, which occupies a fixed level, depends on the biological species. The
universal form of the function $f(n)$ strongly suggests the presence of a
bias of very general origin. Indeed, in \cite{MiSc}, a mutation model has
been proposed where the intensity of the mutation matrix, essentially in
one point mutation, depends from the variation of the labels identifying
the states of the irrep. of $U_{q \to 0}(sl(2) \oplus sl(2))$ describing,
in the crystal basis \cite{FSS}, the codons. The numerically computed
stationary solution of the master equation for the distribution of the 64
codons is nicely fitted by a function of the form eq. (\ref{eq:bf}). The
model depends from the choice of the form of the fitness function and from
the values of the arbitrary parameters appearing in the mutation matrix,
but unrealistic choice of these values destroys the goodness of the fit.
The remark of \cite{GZ04} on the {\it mystery of the straight line} in the
codon position-specific frequencies as function of the total exonic GC
content, raises, on the light of the previously discussed form of the rank
ordered distribution, an even more intriguing question: which is the
mechanism which ensures that the sum of the frequencies $f(n)$, given by
eq. (\ref{eq:bf}) over a set of codons, whose rank distribution generally
depends on the biological species, is a linear function in
$y_{\mathrm{GC}}$? From this property, we say that there is a
\emph{conspiracy} between the different codons. These effects are more
evident when we sum the frequencies of the codons with fixed third
nucleotide, which implies constraints on the amino-acids distribution. We
have also discussed in detail the structure of the Shannon entropy. It is
commonly stated in the literature that the non-coding part of DNA exhibits
more correlation than the coding part, which is in contradiction with what
one would naively expect as the coding part is more subjected to functional
constraints. The results of our work points out the existence of strong
correlations in the exonic part, which very likely witness the existence of
functional bias, worthwhile of further analysis, and which should be better
interpreted and understood.

It has been known for some time that the plots of the first/second codon
position GC content versus the third codon position GC content (or versus
the total GC content) show straight line correlations \cite{MuOs87,
Sueoka88, KFL, Fors}. The straight line behaviour of the codon
position-specific frequencies remarked in \cite{GZ04} (see also Tables
\ref{table:lineareu} and \ref{table:lineararchae}) of course implies these
correlations, but is stronger. In \cite{KFL}, a four-parameter model is
developed to explain the 64 codons and the 20 amino-acids usage frequencies
in function of the GC content. Considering the sums of the eight codons
with the same GC content in position (which correspond to the columns of
our codon matrix (\ref{eq:codonmatrix})), a satisfactory agreement between
experimental values and theoretical curves is found. The authors conclude
that the GC content primarily drives the codon usage rather than the
inverse. Our results, concerning the GC dependence of the codon usage,
although settled on different grounds, go in the same direction.

As a last remark, let us make some comments on genes. Although the GC
content varies much less inside a genome than across genomes for bacteria,
it is instructive to see whether the straight line effect also occurs at
the gene level. In the case of \textit{E. coli K12} for example, for which
the mean GC content is 51.8\%, most of the genes correspond to values of GC
close to the mean value, but the tails of the $P_X^i$ distributions ($X =
\mathrm{C, G, U, A}$ and $i=1,2,3$) show a trend towards straight lines.
Surely, a more detailed study deserves attention. Analogous trend for
\textit{Homo sapiens} has been observed in \cite{Fors}.

Finally we have conjectured the existence of a quasi-symmetry with respect
to the principal and secondary diagonal of the codon matrix, written in the
form suggested by the crystal basis model of the genetic code \cite{FSS}.
Clearly these discrete symmetries can be formulated without any reference
to the crystal basis, but they appear naturally in this mathematical
modelisation. So, as conclusive remark, let us point out that crystal basis
model seems to provide the \emph{kinematical variables} useful to describe
some properties of the genetic code. As it is well known, the use of
appropriate variables in mathematics, in physics, and, very likely, also in
biology is an essential step for facing effectively a problem. 

\bigskip

\textbf{Acknowledgements:} L.F. would like to thank the INFN of Naples and
A.S. would like to thank Universit\'e de Savoie, both for partial financial
supports.

\clearpage

\begin{table}[htbp]
\caption{Regression coefficients for $\sum P(X_{i}) = a \, y_{\mathrm{GC}}
+ b$ (eubacteria).
\label{table:lineareu}}
\centering
\vspace*{12pt}
\begin{tabular}{|c|c|c|c|}
   \hline
   $\sum P(X_{i})$ & $a$ & $b$ & $R$ (Pearson) \\ 
   \hline
   $\sum P(\mathrm{C}NN')$ & $0.424$ & $0.004$ & $0.949$ \\
   $\sum P(\mathrm{A}NN')$ & $-0.452$ & $0.490$ & $0.960$ \\
   $\sum P(\mathrm{G}NN')$ & $0.283$ & $0.209$ & $0.928$ \\
   $\sum P(\mathrm{U}NN')$ & $-0.255$ & $0.297$ & $0.955$ \\
   $\sum P(N\mathrm{C}N')$ & $0.239$ & $0.108$ & $0.925$ \\
   $\sum P(N\mathrm{A}N')$ & $-0.338$ & $0.467$ & $0.959$ \\
   $\sum P(N\mathrm{G}N')$ & $0.215$ & $0.065$ & $0.955$ \\
   $\sum P(N\mathrm{U}N')$ & $-0.117$ & $0.358$ & $0.898$ \\
   $\sum P(NN'\mathrm{C})$ & $1.067$ & $-0.257$ & $0.984$ \\
   $\sum P(NN'\mathrm{A})$ & $-0.928$ & $0.676$ & $0.985$ \\
   $\sum P(NN'\mathrm{G})$ & $0.770$ & $-0.130$ & $0.984$ \\
   $\sum P(NN'\mathrm{U})$ & $-0.909$ & $0.711$ & $0.977$ \\
   \hline
\end{tabular}
\end{table}

\begin{table}[htbp]
\caption{Regression coefficients for $\sum P(X_{i}) = a \, y_{\mathrm{GC}}
+ b$ (archaea).
\label{table:lineararchae}}
\centering
\vspace*{12pt}
\begin{tabular}{|c|c|c|c|}
   \hline
   $\sum P(X_{i})$ & $a$ & $b$ & $R$ (Pearson) \\ 
   \hline
   $\sum P(\mathrm{C}NN')$ & $0.414$ & $-0.028$ & $0.951$ \\
   $\sum P(\mathrm{A}NN')$ & $-0.557$ & $0.569$ & $0.963$ \\
   $\sum P(\mathrm{G}NN')$ & $0.393$ & $0.173$ & $0.917$ \\
   $\sum P(\mathrm{U}NN')$ & $-0.249$ & $0.286$ & $0.901$ \\
   $\sum P(N\mathrm{C}N')$ & $0.269$ & $0.079$ & $0.891$ \\
   $\sum P(N\mathrm{A}N')$ & $-0.285$ & $0.440$ & $0.841$ \\
   $\sum P(N\mathrm{G}N')$ & $0.190$ & $0.082$ & $0.859$ \\
   $\sum P(N\mathrm{U}N')$ & $-0.174$ & $0.398$ & $0.845$ \\
   $\sum P(NN'\mathrm{C})$ & $1.047$ & $-0.241$ & $0.982$ \\
   $\sum P(NN'\mathrm{A})$ & $-0.923$ & $0.687$ & $0.988$ \\
   $\sum P(NN'\mathrm{G})$ & $0.687$ & $-0.066$ & $0.943$ \\
   $\sum P(NN'\mathrm{U})$ & $-0.812$ & $0.619$ & $0.983$ \\
   \hline
\end{tabular}
\end{table}

\clearpage

\begin{table}[htbp]
\caption{Sum of frequencies for drawing lots of $n$ codons belonging to some 
subset $I$ of size $N$. 
\label{table:sort}}
\centering
\vspace*{12pt}
\begin{tabular}{|c|c|c|c|c|c|c|}
    \hline
    sum & $n$ & $N$ & $\overline{a}$ & $\sigma_{a}$ & $\overline{b}$ & $\sigma_{b}$ \\ 
    \hline
    $\displaystyle\sum_{N,N'}P(\mathrm{C}NN')+P(\overline{\mathrm{C}NN'})$
    & 4 & 1800 & $-0.007$ & $0.107$ & $0.123$ & $0.055$ \\
    idem & 8 & 4000 & $-0.013$ & $0.12$ & $0.247$ & $0.064$ \\
    idem & 12 & 1800 & $-0.02$ & $0.107$ & $0.370$ & $0.055$ \\
    \hline
    $\displaystyle\sum_{N,N'}P(N\mathrm{C}N')+P(\overline{N\mathrm{C}N'})$
    & 4 & 1800 & $-0.024$ & $0.113$ & $0.144$ & $0.060$ \\
    idem & 8 & 4000 & $-0.048$ & $0.13$ & $0.287$ & $0.07$ \\
    idem & 12 & 1800 & $-0.074$ & $0.11$ & $0.432$ & $0.06$ \\
    \hline
    $\displaystyle\sum_{N,N'}P(NN'\mathrm{C})+P(\overline{NN'\mathrm{C}})$
    & 4 & 1800 & $0.034$ & $0.116$ & $0.105$ & $0.065$ \\
    idem & 8 & 4000 & $0.067$ & $0.134$ & $0.21$ & $0.074$ \\
    idem & 12 & 1800 & $0.10$ & $0.117$ & $0.314$ & $0.065$ \\  
    \hline
    $\displaystyle\sum_{N,N'}P(\mathrm{G}NN')+P(\overline{\mathrm{G}NN'})$
    & 4 & 1800 & $0.007$ & $0.115$ & $0.126$ & $0.061$ \\
    idem & 8 & 4000 & $0.013$ & $0.132$ & $0.253$ & $0.070$ \\
    idem & 12 & 1800 & $0.021$ & $0.115$ & $0.379$ & $0.061$ \\
    \hline
    $\displaystyle\sum_{N,N'}P(N\mathrm{G}N')+P(\overline{N\mathrm{G}N'})$
    & 4 & 1800 & $0.025$ & $0.108$ & $0.106$ & $0.035$ \\
    idem & 8 & 4000 & $0.050$ & $0.124$ & $0.211$ & $0.063$ \\
    idem & 12 & 1800 & $0.074$ & $0.108$ & $0.318$ & $0.055$ \\
    \hline
    $\displaystyle\sum_{N,N'}P(NN'\mathrm{G})+P(\overline{NN'\mathrm{G}})$
    & 4 & 1800 & $-0.035$ & $0.103$ & $0.145$ & $0.049$ \\
    idem & 8 & 4000 & $-0.074$ & $0.120$ & $0.293$ & $0.057$ \\
    idem & 12 & 1800 & $-0.104$ & $0.103$ & $0.436$ & $0.049$ \\
    \hline
    $\displaystyle\sum_{N,N',N''}P(NN'N'')+P(\overline{NN'N''})$ & 16 &
    10000 & $-0.0024$ & $0.21$ & $0.50$ & $0.112$ \\
    $\displaystyle\sum_{N,N',N''}P(NN'N'')+P(\overline{NN'N''})$ & 24 &
    10000 & $-0.0042$ & $0.245$ & $0.75$ & $0.128$ \\
    $\displaystyle\sum_{N,N',N''}P(NN'N'')$ & 32 & 10000 & $8\;10^{-4}$ &
    $0.30$ & $0.50$ & $0.147$ \\
    \hline
    $\displaystyle\sum_{N,N',N'' \in I_{s}}P(NN'N'')$ & $16$ & 10000 &
    $0.45$ & $0.20$ & $0.019$ & $0.088$ \\
    $\displaystyle\sum_{N,N',N'' \in I_{s}}P(\overline{NN'N''})$ & $16$ &
    10000 & $-0.445$ & $0.189$ & $0.48$ & $0.106$ \\
    $\displaystyle\sum_{N,N',N'' \in I_{s}}P(NN'N'')+P(\overline{NN'N''})$
    & $16$ & 10000 & $0.002$ & $0.18$ & $0.50$ & $0.094$ \\
    \hline
    $\displaystyle\sum_{N,N',N'' \in I_{d}}P(NN'N'')$ & $16$ & 10000 &
    $-0.232$ & $0.189$ & $0.328$ & $0.098$ \\
    $\displaystyle\sum_{N,N',N'' \in I_{d}}P(\overline{NN'N''})$ & $16$ &
    10000 & $0.229$ & $0.226$ & $0.173$ & $0.113$ \\
    $\displaystyle\sum_{N,N',N'' \in I_{d}}P(NN'N'')+P(\overline{NN'N''})$
    & $16$ & 10000 & $-0.003$ & $0.18$ & $0.50$ & $0.094$ \\
    \hline
\end{tabular}
\end{table}

\end{document}